\documentclass[aps,prl,twocolumn,floatfix,amsmath]{revtex4-1}

\usepackage{bm}
\usepackage{graphicx}
\usepackage{latexsym}
\usepackage{amssymb}

\DeclareMathOperator*{\SumInt}{%
\mathchoice%
 {\ooalign{$\displaystyle\sum$\cr\hidewidth$\displaystyle\int$\hidewidth\cr}}
 {\ooalign{\raisebox{.14\height}{\scalebox{.7}{$\textstyle\sum$}}\cr\hidewidth$\textstyle\int$\hidewidth\cr}}
 {\ooalign{\raisebox{.2\height}{\scalebox{.6}{$\scriptstyle\sum$}}\cr$\scriptstyle\int$\cr}}
 {\ooalign{\raisebox{.2\height}{\scalebox{.6}{$\scriptstyle\sum$}}\cr$\scriptstyle\int$\cr}}
}
\def\Fig#1{Fig.\,\ref{#1}}
\renewcommand{\vec}[1]{\mathbf{#1}}
\def\d{\mathrm{d}}

\usepackage{color}
\definecolor{blue}{rgb}{0,0,0.9}\def\jm#1{{\color{black}#1}}
\definecolor{red}{rgb}{0.9,0,0}
\definecolor{green}{rgb}{0,0.8,0}

\begin{document}

\title{\jm{Electron dynamics driven by light pulse derivatives}}

\author{Qi-Cheng Ning, Ulf Saalmann, and Jan M. Rost}
\affiliation{Max Planck Institute for the Physics of Complex Systems, N\"othnitzer Stra{\ss}e 38, 01187 Dresden, Germany}

\date{\today}

\begin{abstract} We demonstrate that ultrashort pulses carry the possibility for
a new regime of light-matter interaction with nonadiabatic electron processes
sensitive to the envelope-derivative of the light pulse.  A standard 
single pulse with its two peaks in the derivative separated by the width of the
 pulse acts in this regime like a traditional double pulse. The two
ensuing nonadiabatic ionization bursts have slightly different ionization
amplitudes. This difference is due to redistribution of continuum electron energy
during the bursts, negligible in standard photo-ionization. A time-dependent
close-coupling approach based on  cycle-averaged potentials in the Kramers-Henneberger reference frame permits a detailed understanding of light pulse derivative-driven electron dynamics.
\end{abstract}

\maketitle 

With increasingly shorter pulses becoming available in the optical and VUV
domain, it has been noticed that the light pulse envelope influences the outcome
of experiments. Most widely known is the influence of the carrier envelope phase
quantifying the shift of the periodic carrier wave relative to the maximum of
the pulse envelope \cite{pagr+01}. Also, effects of the spatial envelope
dependence beyond the dipole approximation of light-matter coupling
\cite{fose+05,sikj+16} and a chirp on the ultrashort pulse \cite{shro16} have
been pointed out. A more indirect effect is induced by the time-dependent
AC-Stark shift of energy levels which follows the pulse envelope. It gives rise
to dynamic interference \cite{toto+07,toto+08}, termed as such and elegantly
explained in \cite{dece12,dece13}.
Inspired by earlier work \cite{toto+07}, the envelope Hamiltonian was introduced
\cite{tosa+15}, which explicitly separates the optical periodic time dependence
from that of the envelope variation and permits therefore a clear distinction
of multiphoton \cite{mesi99} and nonadiabatic \cite{toto+09} ionization under short pulses. Much earlier
and on much longer time scales of Rydberg excitation and femtosecond pulses, pulse enevelope effects, mostly
in connection with transient Stark shift enabled resonances,
were pointed out  \cite{frbu+87,stga93,stdu+93,stdu+94,jo95,scku97}.

Despite these various notions on effects of the pulse envelope the simple but
dramatic consequences for nonadiabatic ionization and the possibilities these
consequences carry have not yet been addressed: Plainly put, in the regime of extreme
 nonadiabatic matter-light coupling, electron dynamics becomes sensitive to the light-pulse derivatives. Hence, a standard Gaussian laser pulse acts as a double pulse through the two peaks of its
derivative. \jm{With the Gaussian pulse as an example we will establish that in general nonadiabatic electron dynamics is sensitive to the
pulse envelope derivative (PED) of the laser pulse   rather than to the envelope (maximum) itself}.

nonadiabatic dynamics  occurs for states which change fast as a function of an external parameter \cite{na12}. 
Molecules are the most widely known examples, where the electronic Born-Oppenheimer states  depend parametrically on the nuclear positions. In the present context, we formulate the electronic state as parametrically dependent
on the pulse envelope. \jm{The  ensuing 
nonadiabatic ionization is exclusively due to
 PED, as we will see below.} Therefore, nonadiabatic ionization is complementary to dynamic interference,
resonant population trapping or Rydberg multiphoton ionization \cite{dece12,frbu+87,stga93,stdu+93,stdu+94,jo95,scku97}. These  are adiabatic phenomena in the sense that a resonance condition for an energy difference of bound states, well
defined at each time during the pulse, is fulfilled twice,
during the rise and fall of the pulse, respectively. \jm{These two times are in
general not where the envelope derivative peaks, as is the case for nonadiabatic ionization. }
Moreover, the resonant effects mask  nonadiabatic ionization and its low energy photo electron peak as we discuss here. 
Otherwise, nonadiabatic ionization 
could have already been identified in the 1990s, in particular in Rydberg experiments such as \cite{jo95}.
 
To demonstrate that nonadiabatic ionization is indeed only sensitive to the
{\em change} of the pulse envelope, we will consider a pulse with a short rise
and fall encompassing a plateau of variable duration $T_{\rm c}$. This allows us to
analyze and understand the subtle differences of the beginning and the end of
the pulse separately and to demonstrate that illumination with maximal amplitude
during the plateau has no effect on nonadiabatic ionization. Shrinking the
plateau to $T_{\rm c} = 0$, we will arrive at the normal single ultrashort pulse,
whose effect is then easily understood in terms of the (already analyzed) rising
and falling part of the pulse.

\begin{figure*}[tb]
\centering
\includegraphics[width=\textwidth]{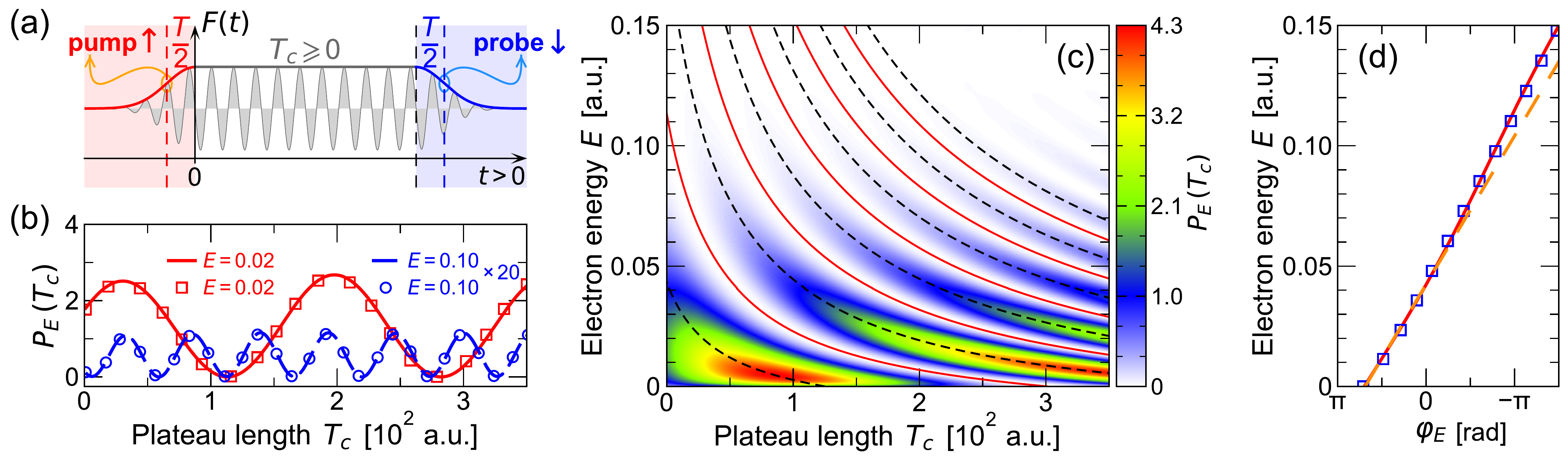}
\caption{(color online).
nonadiabatic ionization.
Panel (a): A flattop laser pulse rising $(\uparrow$) and falling ($\downarrow$)
over a time span of $T/2=25.5\,$au\ between the maximum of the electric field amplitude $F(t)$ and its maximal derivative.
These two Gaussian half-pulses, which act similarly as a pump and probe pulse in the nonadiabatic regime, encompass a plateau of duration $T_{\rm c}$. Panel (b): Electron spectra
for two energies as a function of plateau length $T_{\rm c}$ for an electron
initially bound by a weak potential at energy $E_{\rm g} = -0.0277\,$ a.u.\ exposed to a laser
pulse of peak amplitude $F = 0.5\,$ a.u.\ and frequency $\omega=0.314\,$ a.u., see text. TDSE solutions
with the full Hamiltonian are given as lines and with $H_{0}$ from
Eq.\,\eqref{eq:TISE} as symbols. Panel (c): nonadiabatic electron spectrum
$P_{E}(T_{\rm c})$. The lines mark maxima and minima predicted from
Eq.\,\eqref{eq:pump-probe_maxmin}. Panel (d): Phase difference $\varphi_{E}$,
extracted by fitting Eq.\,\eqref{eq:pump-probe} to spectra for fixed $E$ as in
\Fig{Fig1}b (symbols); from benchmark calculations obtained through
wave packet partitioning (see text) for the Gaussian pulse, i.\,e.\ $T_{\rm c}=0$ (solid
line); approximated with Eq.\,\eqref{eq:phase} (dashed line).
}
\label{Fig1}
\end{figure*}

Although the envelope-derivative effects we are going to investigate are
independent of the theoretical description, we deliberately choose the envelope
Hamiltonian \cite{tosa+15}, since it reveals in connection with a time-dependent
close-coupling (CC) representation the mechanism of envelope-driven nonadiabatic electron
dynamics, including subtle effects such as the reshuffling of electron energy in
the continuum. To this end we Fourier-expand the periodic time dependence of the
electron potential $V(\vec x,t)$ in the Kramers-Henneberger frame, keeping the
time evolution of the pulse envelope explicit (atomic units (au) are used unless
stated otherwise),
\begin{equation}\label{EnvelopePot}
V_n(\vec x,t)=\frac{\omega}{2\pi}\int_0^{\frac{2\pi}{\omega}}\!\!\!\!\!{\rm e}^{{\rm i} n\omega t'} V\big(\vec x+\alpha(t)\,\vec{e}_{z}\cos(\omega t'+\varphi)\big)\, \d t',
\end{equation}
where $\vec x_{\omega}(t)=\alpha(t)\,\vec{e}_{z}\cos(\omega t+\varphi)$ can be understood as the trajectory of a free electron in the laser electrical field linearly polarized along $\vec{e}_{z}$ and defined by the 2nd derivative of $\vec x_{\omega}(t)$. While to a very good accuracy an expansion length of $n_\mathrm{max}=2$ is sufficient in the potentials \eqref{EnvelopePot} as shown before \cite{tosa+15}, nonadiabatic dynamics sensitive to the envelope derivative $\d\alpha/\d t$, is mainly described through $V_{0}(\vec x,t)$ with eigenstates and energies parametrically dependent on time in
\begin{equation}\label{eq:TISE}
[H_0(t)-E_{\beta}(t)]|\psi_{\beta}(t)\rangle= 0\,,
\end{equation}
where $H_{0}=-\frac 12 \bm\nabla^{2}+V_0$.

This is directly illustrated for a flattop pulse in \Fig{Fig1}. It is constructed from a Gaussian pulse extended by inserting at its maximum a plateau of length $T_{\rm c}$, see \Fig{Fig1}a. To keep the analysis as simple as possible, we use a one-dimensional model potential for a weakly bound electron introduced in different contexts before \cite{poti+99} as a specific example. We compare the photoelectron spectrum
obtained from the solution of the time-dependent Schr\"odinger equation (TDSE) with the full Hamiltonian and with $H_{0}$. Both solutions agree quite well for slow electrons $E/\omega \ll 1$ (compare lines and symbols in \Fig{Fig1}b) implying a nonadiabatic regime with envelope-derivative driven electron dynamics. Since this derivative has two peaks during the rise and fall of the pulse, respectively, but vanishes during the plateau of the pulse, we expect two ionization bursts which generate a typical two-slit interference pattern as a function of plateau length $T_{\rm c}$ in the electron spectrum,
\begin{align}\label{eq:pump-probe}
P_{E}(T_{\rm c})&=a_E\cos\left(\varphi_{E}-\delta_{E}T_{\rm c}\right)+c_E\,.
\end{align}
Indeed, for any fixed energy $E$ the ionization yield oscillates perfectly as a function of plateau length $T_{\rm c}$ as shown in \Fig{Fig1}b. Fitting \eqref{eq:pump-probe} to these yields we can extract $\varphi_{E}$ and $\delta_{E}$,
which allows us to determine from
\begin{equation}\label{eq:pump-probe_maxmin}
\varphi_{E}-\delta_{E}T_{\rm c}=n\pi,
\end{equation}
the maxima ($n$ even, dashed) and minima ($n$ odd, solid) in very good agreement
with the numerical spectra as shown in \Fig{Fig1}c.
 In the $(E,T_{\rm c})$ plane, the functional form $T_{\rm c}\propto E^{-1}$ of these
 extrema follows directly from the difference between the final and initial
 energy, $\delta_{E}=E-E_{\rm g}^{\star}$. Note however, that in contrast to
 standard double pulses, the light pulse illuminates the target with maximal
 amplitude between the (nonadiabatic) ionization bursts. Therefore,
 $E_{\rm g}^{\star}$ is the initial energy dressed by the laser field, as indicated
 by the star.

\begin{figure}[tph]
\centering
\includegraphics[width=\linewidth]{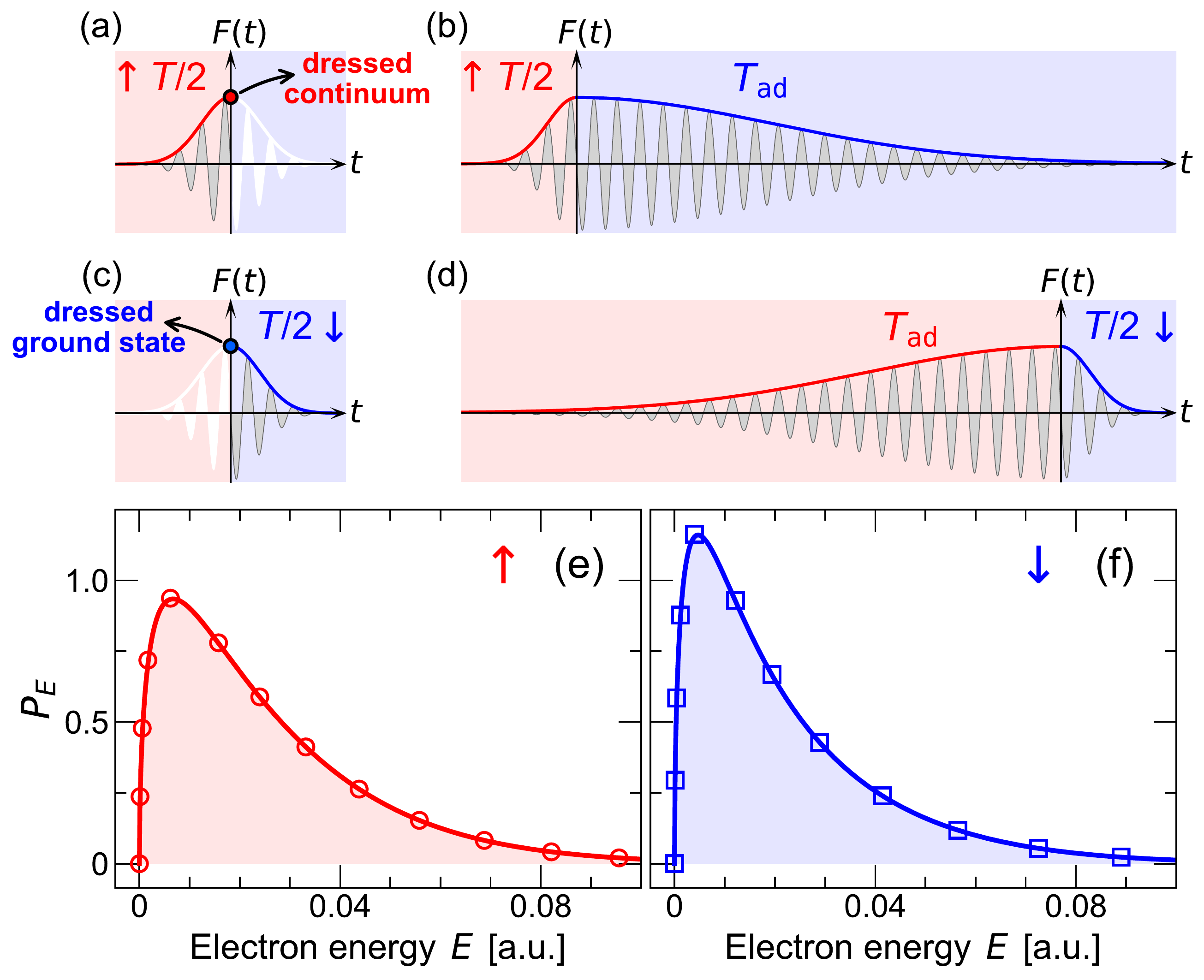}
\caption{(color online).
nonadiabatic electron spectra from Gaussian half-pulses. (a) The sketch of the
partitioning approach for the rising ($\uparrow$) half-pulse. (b) The artificial
laser field with the falling pulse length $T_\mathrm{ad}$ long enough to not
affect nonadiabatic dynamics while the rising part has as before $T/2=25.5\,$au.
(Note that for the figure sketch $T_\mathrm{ad}=170\,$ a.u.\ while in the calculations
$T_\mathrm{ad}=850\,$ a.u.\ has been used, which is sufficiently long to obtain
converged results). Panel (e): The photoelectron spectrum from the partitioning
approach as sketched in (a) (lines) and from the artificial laser field as in
(b) (symbols). The panels (c), (d) and (f) show information analogous to (a),
(b) and (e), respectively, but for the falling ($\downarrow$) half-pulse instead
of the rising half-pulse. Peak field strength and laser frequency are the same as in \Fig{Fig1}.}
\label{Fig2}
\end{figure}

As a next step we take a closer look at the ionization bursts themselves. The
fast rising and falling half-pulses generate the electron spectra shown in
\Fig{Fig2}e and \ref{Fig2}f with solid lines, respectively.
We have obtained these spectra by {\it wave packet partitioning}: We solve the
TDSE for $H_{0}$ with the electron initially in the ground state
$\psi_{\rm g}(t\to-\infty)$. Projecting at the end of the rising half-pulse at $t=0$
onto the instantaneous continuum eigenstates $\psi_{E}$ of $H_{0}$ at maximal
field (\Fig{Fig2}a), we obtain the spectrum of the 1st burst (\Fig{Fig2}e).
For the falling half-pulse we begin the propagation in the (dressed) ground
state $\psi_{\rm g}^{\star}$ at maximal field (\Fig{Fig2}c) and obtain the spectrum
of the 2nd burst from $\psi_{E}(t\to\infty)$, see \Fig{Fig2}f. Underscoring
again the nonadiabatic dynamics, the result is the same (see solid lines in
Figs.\,\ref{Fig2}e,f) if we amend the left half-pulse with a slowly decaying tail (half
width $T_\mathrm{ad}=850\,$ a.u., \Fig{Fig2}b) and start the right half-pulse with an equally
slow rise ($T_\mathrm{ad}=850\,$ a.u., \Fig{Fig2}d). Since the concatenation of the two
half-pulses should be identical to  our pulse
pulse for $T_{\rm c}=0$, i.\,e., a Gaussian pulse, we expect that the phase
$\varphi_{E}$ in \eqref{eq:pump-probe} is given by the phase difference
$\varphi_{E}=\varphi_{\uparrow}(E)-\varphi_{\downarrow}(E)$ of the two burst
amplitudes $A_{\uparrow\downarrow}(E){\rm e}^{{\rm i}\varphi_{\uparrow\!\downarrow}(E)}$ .
This is indeed the case, as shown in \Fig{Fig1}d. Note that, suitable for nonadiabatic dynamics, we measure the amplitude pulse length $T$ here as the time span between the maxima in the derivative of the pulse envelope which is related to the standard measure of full width at half maximum (FWHM) of the envelope $\tau$ through $T= \tau/(2\ln 2)^{1/2} = 0.85\,\tau$. 
%

Interestingly, the electron spectra from the rising ($\uparrow$) and falling ($\downarrow$) 
half-pulse differ slightly, although the total (energy-integrated) ionization yield is the same, 
in our example $P_{\uparrow}\equiv \int A_{\uparrow}{\!}^{2}(E)\,\d E = P_{\downarrow}\equiv \int A_{\downarrow}{\!}^{2}(E)\,\d E=0.03247\,$au
\footnote{
One can show in general that two pulses $f(t)$ and $f(-t)$ induce the same depletion of the initial state. Since the system considered here has just one bound state and $F_{\uparrow}(t)=F_{\downarrow}(-t)$ (see, e.g., \Fig{Fig2}), this implies $P_{\uparrow}=P_{\downarrow}$.
}.
This suggests that absorption from the initial state does not depend on the
character of the half-pulse (rising or falling), while there must be a
mechanism of redistributing energy in the continuum, sensitive to the sign of
the pulse derivative. In contrast to the fully numerical solutions presented
so far, a CC representation in a basis allows us to distinguish
nonadiabatic bound-continuum $M_{E\rm g}$ from continuum-continuum $M_{EE'}$ transitions,
with the transition matrix elements
\begin{equation}\label{eq:M_gc}
M_{E\beta}(t)=\frac{\langle \psi_{E}(t)|\partial_t V_0(\vec x,t)|\psi_{\beta}(t)\rangle}{E-E_\beta(t)}\,
\end{equation}
for both cases.
Inserting the wavefunction $\big|\Psi(t)\big>=\SumInt_{\beta} \big|\psi_{\beta}(t)\big>
\,{\rm e}^{-{\rm i}\!\int^{t}\!E_{\beta}(t')\,\d t'}c_{\beta}(t)$
into the TDSE with the Hamiltonian $H_{0}(t)$ the matrix
elements \eqref{eq:M_gc} govern the evolution of the time-dependent amplitudes
$c_{\beta}(t)$ through \cite{tosa+15}
\begin{subequations}
\label{eq:CC}
\begin{align}
\frac{\d c_{\rm g}}{\d t\,\,}&=-\int_{0}^{\infty}c_E M_{E\rm g}\,{\rm e}^{-{\rm i}\phi_{\rm g}}\,\d E\\
\frac{\d c_E}{\d t\,\,}&=c_{\rm g}M_{E\rm g}\,{\rm e}^{{\rm i}\phi_{\rm g}}+\int_0^{\infty}\!\!\!\!\!c_{E'}M_{EE'}{\rm e}^{{\rm i}\phi_{E'}}\,\d E'
\end{align}
with the phases
\begin{equation}\label{eq:phaset}
\phi_{\beta}(t)=\int^{t}\big[E-E_\beta(t')\big]\,\d t'
\end{equation}
\end{subequations}
for the ground state $\beta\,{=}\,\rm g$ or a continuum state with energy $\beta\,{=}\,E'$,
respectively. Note that $M_{E\beta}$ is real and $M_{EE}=0$.

\begin{figure}[tb]
\centering
\includegraphics[width=\linewidth]{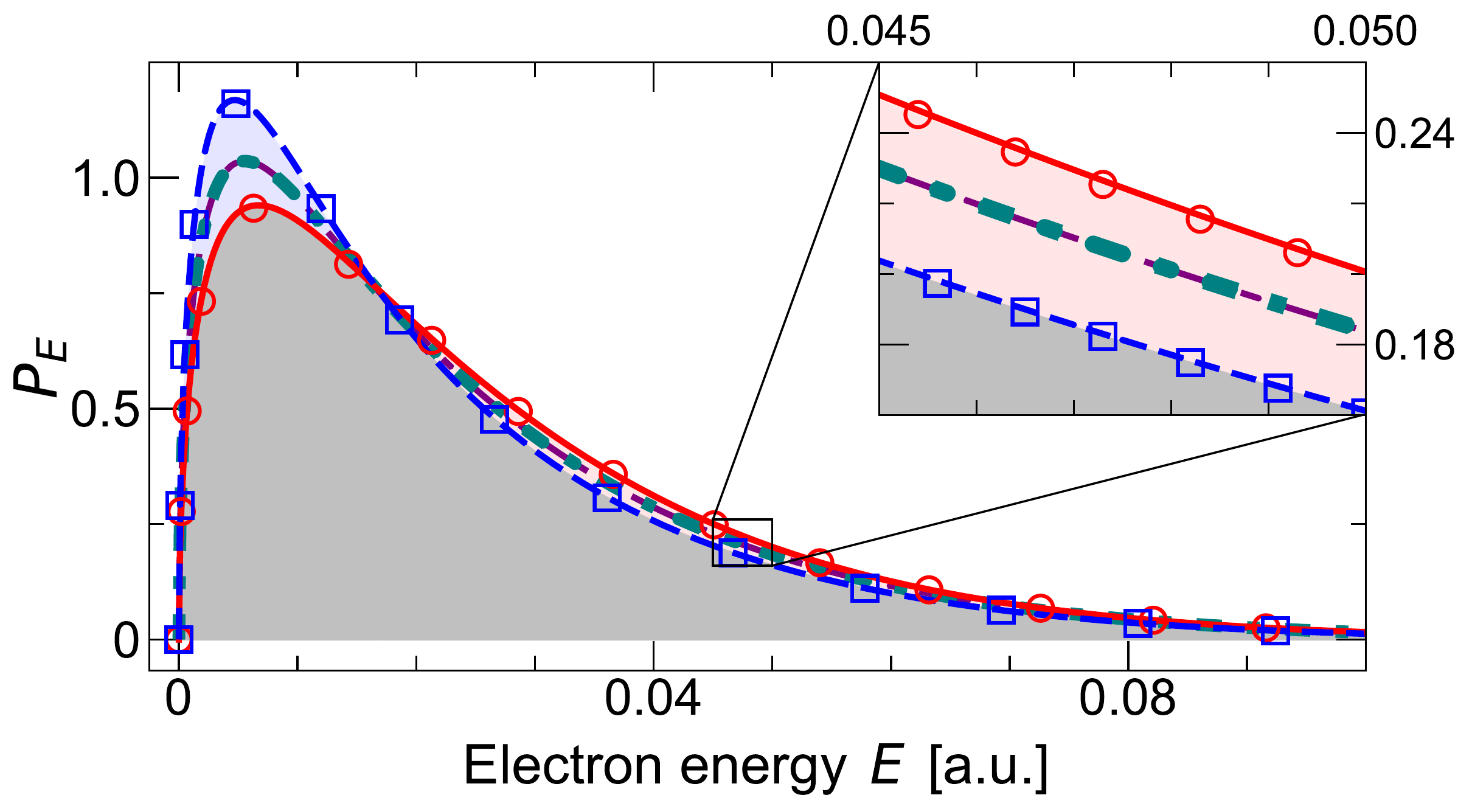}
\caption{(color online).
Energy reshuffling of continuum electron wave packets.
photoelectron spectra from the rising (red) and falling (blue) half-pulses. Results are obtained by the CC equations \eqref{eq:CC} with (solid and dashed lines) and without (dashed-dotted lines) the continuum coupling contribution $M_{EE'}$, as well as with wave packet partitioning (symbols). The laser parameters are the same as in \Fig{Fig1}.}
\label{Fig3}
\end{figure}

As expected, the CC spectra  for the left and right half-pulses are indistinguishable
from our full numerical spectra  obtained by wave packet partitioning, see Fig.\,\ref{Fig3}. 
If we, however, calculate the
left and the right spectra without the continuum-continuum coupling $M_{EE'}$ in
Eqs.\,\eqref{eq:CC}, they become identical, see the dashed-dotted curves in \Fig{Fig3}
and still produce the same total ionization yield as in the full calculation.
Consequently, the already ionized (continuum) electrons are redistributed
towards higher energy through $M_{EE'}$ during the rising half-pulse while being reshuffled towards lower energy in
the continuum for the falling half-pulse.
This explains the difference in the two burst spectra. \jm{The {\it continuum energy
reshuffling} is another (subtle) effect of PED induced
electron dynamics.} It is absent in traditional double pulses with slowly varying
envelope since for those pulses the matrix element $M_{EE'}$ will be negligible.

Following our argument so far, the electron spectrum \eqref{eq:pump-probe} for a
single Gaussian pulse ($T_{\rm c}=0$) is composed from the coherent superposition of
the slightly different burst amplitudes. Since they are created close to the
maxima of the envelope derivative at times $\pm T/2$, their phases differ during
the interval $T$ between the two bursts and therefore the phase difference in
Eq.\,\eqref{eq:pump-probe_maxmin} may be approximated using Eq.\,\eqref{eq:phaset} as
\begin{equation}\label{eq:phase}
\varphi_{E}= - ET + \int_{-T/2}^{+T/2}\!\!E_{\rm g}(t)\,\d t+\pi\,,
\end{equation}
where $\pi$ is a consequence of the opposite sign of the two burst amplitudes.
One sees in \Fig{Fig1}d, that Eq.\,\eqref{eq:phase} describes $\varphi_{E}$ well, in
particular for small energies $E$.

\begin{figure}[tb]
\centering
\includegraphics[width=\linewidth]{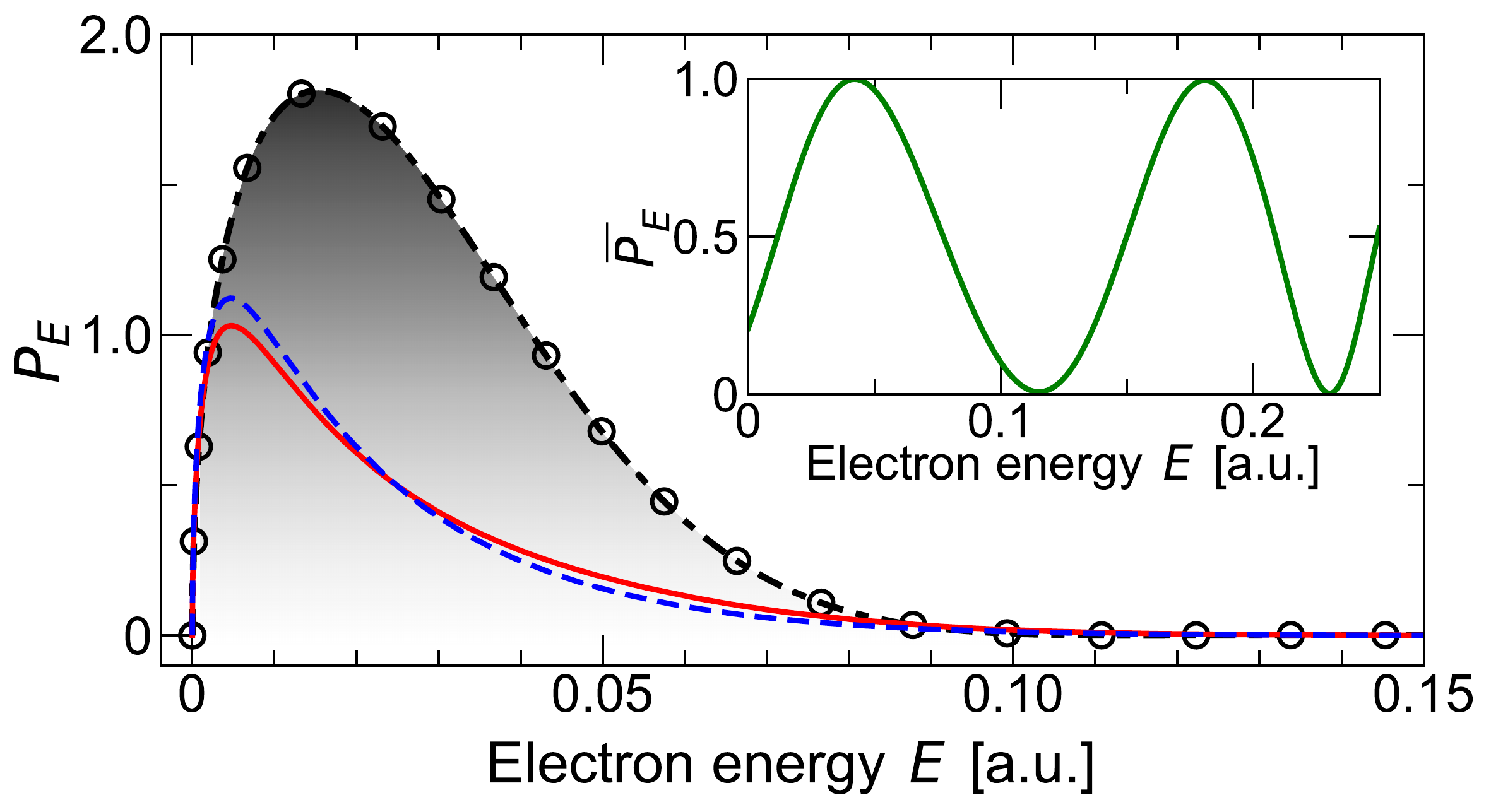}
\caption{(color online).
The nonadiabatic electron spectrum of a single Gaussian pulse of width $T
= 51\,$ a.u.\ (dashed-dotted) and combined from the sequence of a rising and falling
half-pulse with $T/2=25.5\,$ a.u.\ (open circles). In addition the contribution from the
1st electron burst $\tilde A_{\uparrow}{\!}^{2}(E)$ (solid, red) and the 2nd one 
$A_{\downarrow}{\!}^{2}(E)$ (dashed blue) are shown.
The inset reveals Stueckelberg oscillations \cite{st32} of the normalized spectrum, 
$\bar{P}_{E}\equiv P_{E}/[2\tilde A_{\uparrow}{\!}^{2}(E)+2A_{\downarrow}{\!}^{2}(E)]$.
Laser parameters are the same as in \Fig{Fig1}.}
\label{Fig4}
\end{figure}

There is, however, one last element missing, namely that the 1st electron
burst amplitude, $A_{\uparrow}(E)$, gets modified by the 2nd half-pulse to
$\tilde A_{\uparrow}(E)$ in that energy is shuffled through $M_{EE'}$ towards
lower energies, partially canceling the continuum shuffling during the 1st
half-pulse towards higher energies. As a result, the spectra of the two electron
bursts are more similar when combined in a full pulse (blue and red
curve in \Fig{Fig4}) than if considered separately as in \Fig{Fig2} and \Fig{Fig3}. Still, the
two burst amplitudes are not identical after the end of the pulse, apart from a
single point in energy $E \approx 0.02\,$ a.u.\ where they cross. As expected from the
phase difference $\varphi_{E}$ the two burst amplitudes interfere and produce
oscillations in the spectrum as a function of energy $E$. Since their period is larger than 
the energy interval covered by the nonadiabatic ionization peak, it is necessary to 
normalize the spectrum with its major variation in energy in order to uncover the oscillations,  see inset of \Fig{Fig4}. 
Hence, our analysis of nonadiabatic ionization in terms of electron bursts induced by half-pulses has
lead us to a surprising re-interpretation of the photoelectron spectrum at low
energies (grey area in \Fig{Fig4}) including the identification and explanation
of an oscillatory structure, clearly visible in the normalized spectrum $\bar{P}_{E}$.

\begin{figure}[b]
\centering
\includegraphics[width=.8\linewidth]{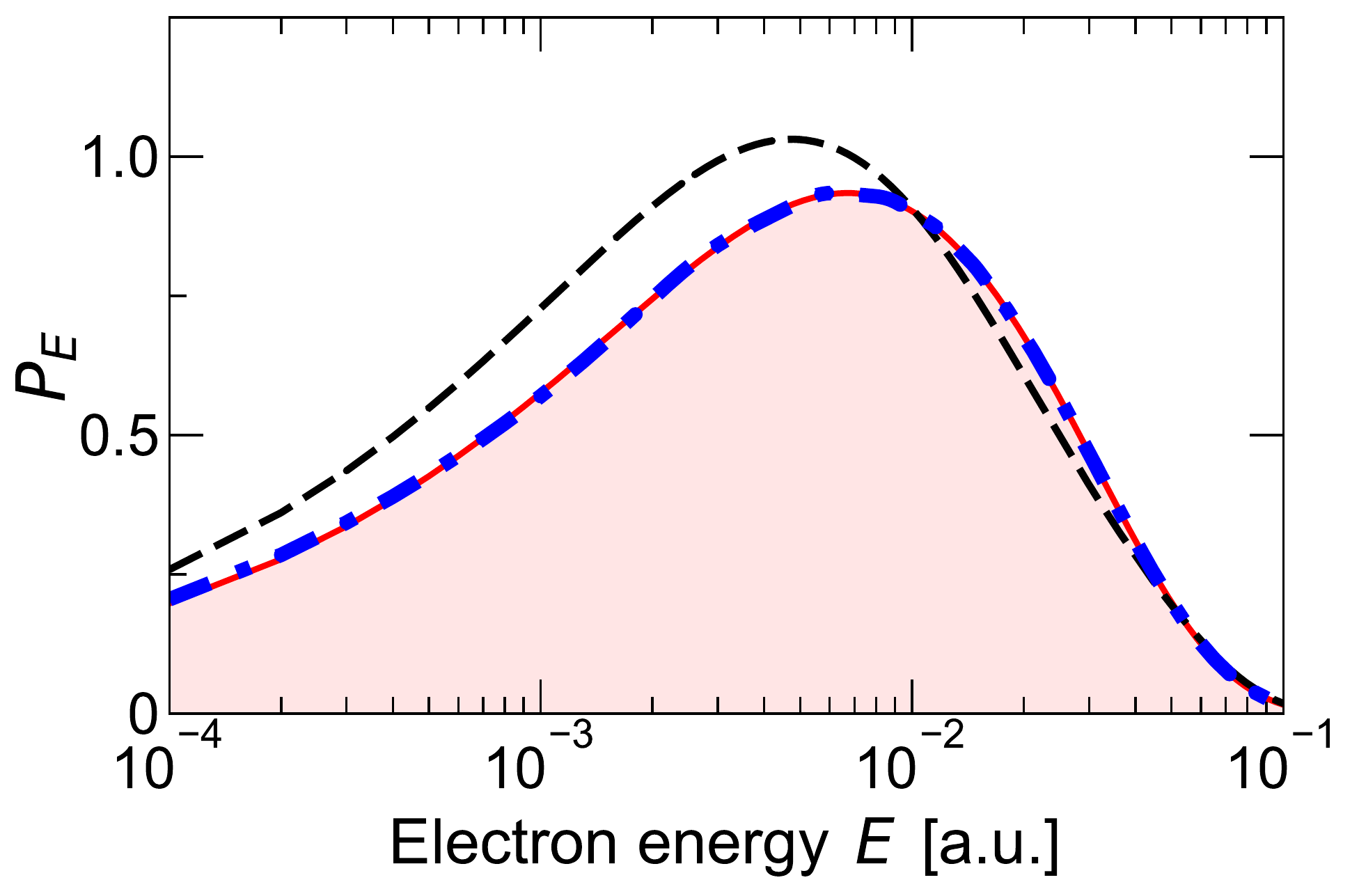}
\caption{(color online).
photoelectron spectrum $A_{\uparrow}{\!}^{2}(E)$ of a rising half-pulse only as in \Fig{Fig2}e (solid red
line) and $\tilde A_{\uparrow}{\!}^{2}(E)$ modified by the falling half-pulse for different plateau lengths
$T_{\rm c}$ ($T_{\rm c}\,{=}\,0$: dashed black, $T_{\rm c}\,{=}\,2000\,$au: dashed-dotted blue). 
Laser parameters are the same as in \Fig{Fig1}.}
\label{Fig5}
\end{figure}

We finally come back to the modification of the 1st burst by the 2nd
half-pulse, which is also known from standard double pulses in the adiabatic
regime. It simply means that the wave packet of the 1st burst is still in the
vicinity of the potential with range $d$ during the 2nd half-pulse.
Modifications are expected if $d/D\equiv d/\sqrt{2E_\mathrm{peak} }/T<1$, where
$D$ is the distance travelled by the continuum wave packet during the time $T$
elapsed between the two bursts, estimated from its most probable energy
$E_\mathrm{peak}$. Note that here $T$ is just the width of the single short (Gaussian)
pulse. With the flattop pulse we have introduced in the beginning, we can
probe the evolution of the modification since there the time between bursts is
given by $T+T_{\rm c}$. Indeed, the modification of the 1st burst vanishes for a
long plateau as one can see in \Fig{Fig5}.

To summarize, in the regime of nonadiabatic light-matter interaction, electron
dynamics is sensitive to the envelope-derivative of a light pulse. Therefore, a
typical short Gaussian pulse acts like a ``double pulse'' through its two maxima in the envelope derivative,
separated by $T= 0.85\,\tau$ where $\tau$ is the FWHM of the Gaussian envelope.
It creates two ionization bursts, which can also be interpreted as being composed from two
ionization paths for each final electron energy $E$. Between the two bursts
electron amplitude of the 1st path is already in the continuum, while the one
of path two is still in the (laser-dressed) ground state. This gives rise to a
phase difference proportional to the energy difference of the two paths and the
time $T$ over which this energy difference exists and leads to an interference
structure in the nonadiabatic part of the electron spectrum produced by a
single short Gaussian pulse. Another subtle feature is the energy reshuffling in the continuum which has the
opposite effect on bursts produced by the rising and falling half-pulse,
respectively.

Clearly, nonadiabatic short-pulse-induced electron dynamics carries unusual
features which we have described here. They can occur,
whenever the pulse envelope changes on the relevant electronic time scale. They will be most prominent for ultrashort pulses, where  resonant excitation is less likely to dominate.   Sensitive to the derivative of the pulse
envelope, these features  \jm{provide new avenues to coherently steer
electron dynamics when light-pulse derivatives can be controlled.}

We thank Koudai Toyota for valuable discussions at an early stage of this project.
This work was supported by the Marie Curie Initial Training Network CORINF and
the DFG priority program QUTIF (SPP 1840).

\end{document}